\documentclass[11pt, a4paper]{article}

\usepackage{amsmath}
\usepackage{amsfonts}
\usepackage{amssymb}
\usepackage{graphicx, rotating}
\usepackage{epstopdf}
\usepackage{epsfig}
\usepackage{latexsym}
\usepackage{graphicx}
\usepackage{color}
\usepackage{amsmath,bm,amssymb}
\usepackage{cite}
\usepackage{slashed}
\usepackage{soul}
\usepackage[hyperfootnotes=true]{hyperref}
\hypersetup{colorlinks, citecolor=bluscuro, linkcolor=black, urlcolor=bluscuro}
\definecolor{bluscuro}{rgb}{0.15, 0.2, .85}
\usepackage[multiple]{footmisc}

\newcommand{\eq}[1]{eq.~(\ref{#1})}

\newcommand{\be}{\begin{equation}}
\newcommand{\ee}{\end{equation}}
\newcommand{\bea}{\begin{eqnarray}}
\newcommand{\eea}{\end{eqnarray}}

\newcommand{\gsim}{\lower.7ex\hbox{$\;\stackrel{\textstyle>}{\sim}\;$}}
\newcommand{\lsim}{\lower.7ex\hbox{$\;\stackrel{\textstyle<}{\sim}\;$}}

\definecolor{nicered}{rgb}{0.7,0.1,0.1}
\definecolor{nicegreen}{rgb}{0.1,0.5,0.1}
\hypersetup{colorlinks,citecolor= nicegreen,linkcolor= nicered}

\begin{document}

\begin{titlepage}
\begin{flushright}
\end{flushright}
\vspace{.3in}

\vspace{1cm}
\begin{center}
{\Large\bf\color{black}
Large tensor-to-scalar ratio and running\\\vspace{0.1cm} of the scalar spectral index with Instep Inflation }
\\
\bigskip\color{black}
\vspace{1cm}{
{\large Guillermo Ballesteros$^a$ and J.\ Alberto Casas$^b$}
\vspace{0.3cm}
} \\[7mm]
$^a$\emph{Institut f\"{u}r Theoretische Physik, Universit\"{a}t Heidelberg.\\ Philosophenweg 16, D-69120 Heidelberg, Germany.}\\\vspace{0.2cm}
$^b$\emph{Instituto de F\'{i}sica Te\'{o}rica, IFT-UAM/CSIC. Cantoblanco 28049, Madrid, Spain.}\vspace{0.3cm}

{\small g.ballesteros@thphys.uni-heidelberg.de\,, alberto.casas@uam.es}

\end{center}
\bigskip

\vspace{.4cm}

\begin{abstract}
If a sizeable tensor-to-scalar ratio $\sim 0.1$ turns out to be detected and a negative running of the scalar spectral index $\sim 10^{-2}$ is significantly required by the data, the vast majority of single field models of inflation will be ruled out. We show that
a flat tree-level effective potential, lifted by radiative corrections and by the imprints of a high energy scale (in the form of non-renormalizable operators) can explain those features and produce enough inflation in the slow-roll regime. 

\end{abstract}
\vfill

\end{titlepage}

\section{Introduction}
The recent measurement of CMB B-modes by the BICEP2 collaboration \cite{Ade:2014xnap} raised much interest and controversy. The fits performed by BICEP2 suggested the detection of a sizeable  tensor-to-scalar ratio, $r\sim{\cal O}(10^{-1})$ \cite{Ade:2014xnap}. However, these data are subject to considerable foreground uncertainty and it was later shown to be premature to interpret them as a  detection of primordial gravitational waves from an early phase of inflationary expansion. Soon after the release of the BICEP2 data, it was pointed out that assuming a dust contribution to the $\mathcal{C}_\ell$ of the BB spectrum of the form $\sim\ell^{-2.3}$ and marginalizing over its amplitude, only an upper limit $r<0.11$ ($95\%$ C.L.) could actually be obtained from BICEP2 and Planck CMB data \cite{Mortonson:2014bja}. It was argued that if the dust foreground were perfectly well known, this limit could be pushed down to $r<0.05$, or a detection of $r$ of order 0.1 could be achieved \cite{Mortonson:2014bja}. A separate analysis using several seemingly consistent methods to estimate the dust foreground amplitude indicated that the signal was likely to be compatible with $r=0$ and a strong polarized dust component \cite{Flauger:2014qra}. Recently, an analysis by the Planck team has shown that those concerns about the foregrounds were indeed well-founded. The Planck collaboration has made available measurements of the thermal emission from diffuse Galactic dust at various frequencies. The extrapolation of these data to 150GHz shows that the measurement of BICEP2 may indeed be entirely due to dust emission, but the uncertainty is still large \cite{Adam:2014bub}. A more definite conclusion will not be reached until a joint detailed analysis of the two experiments is performed \cite{Adam:2014bub}. The possibility exists that there is a primordial component of gravitational waves in the BICEP2 data, albeit lower than it was initially claimed by the collaboration.\footnote{The significance of the result was somewhat tempered from the first arXiv version of the paper \cite{Ade:2014xna} to the final publication of the BICEP2 collaboration \cite{Ade:2014xnap}. The original value $r=0.16^{+0.06}_{-0.05}$ at $68\%$ C.L. obtained by BICEP2, corresponding to a $5.9\,\sigma$ exclusion of $r=0$ \cite{Ade:2014xnap}, came from the subtraction of the most constraining foreground estimation produced by the BICEP2 team, the so called DDM2 model \cite{Ade:2014xnap}. Serious doubts were later cast on its reliability \cite{Flauger:2014qra}, see also \cite{Fla:2014tak, Zald:2014tak} (and \cite{Liu:2014mpa} for another possible source of contamination). The Planck analysis later showed, that the DDM2 model used by BICEP2 very likely underestimated the dust polarization fraction by a factor of $\sim 2$.}  A recent genus topology study of the BICEP2 B-modes points in this direction and towards a tensor-to-scalar ratio of about $0.1$ \cite{Colley:2014nna}.

\begin{figure}[bt] \label{cls}
\centering
\includegraphics[width=0.47\textwidth]{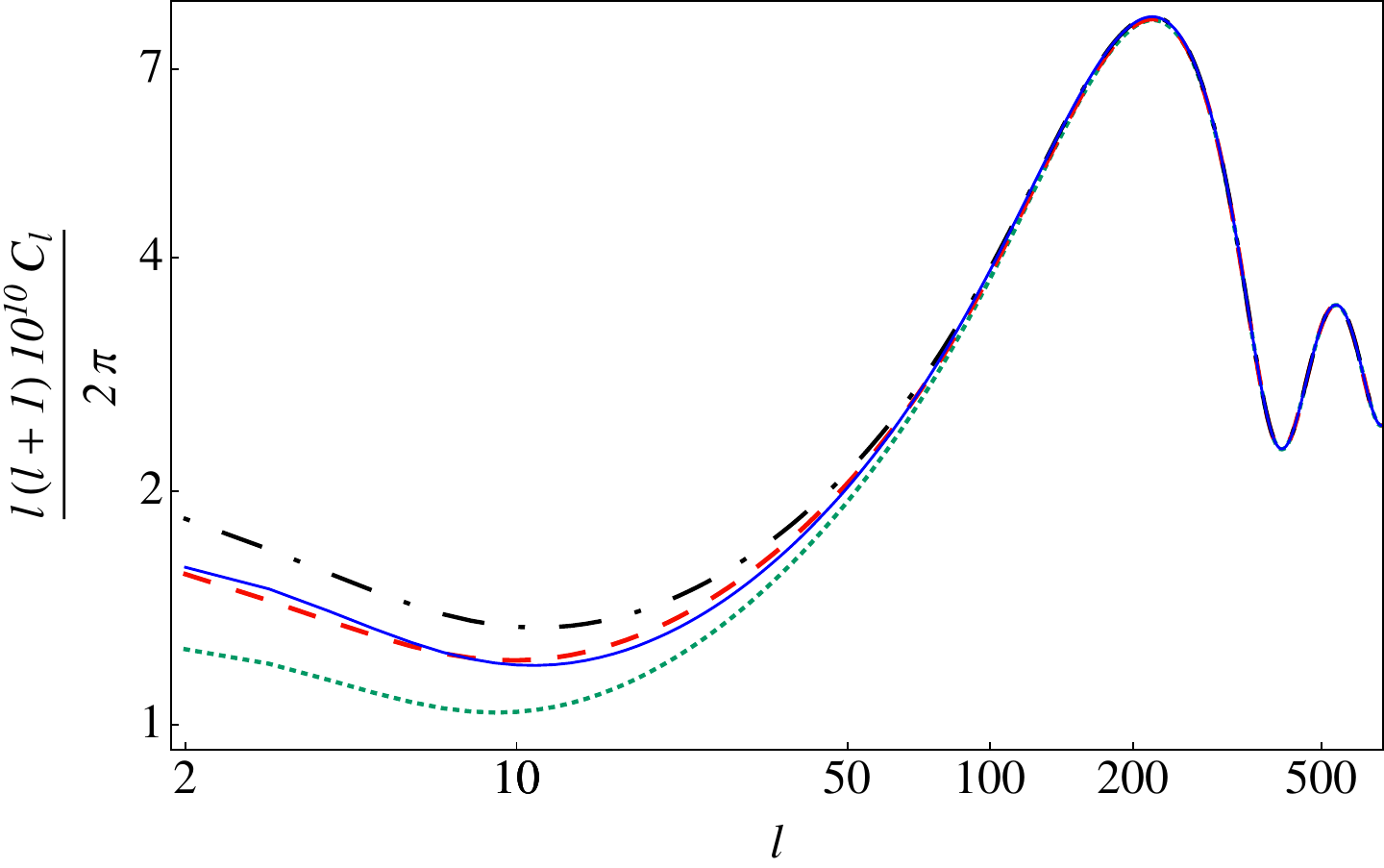}\quad
\includegraphics[width=0.49\textwidth]{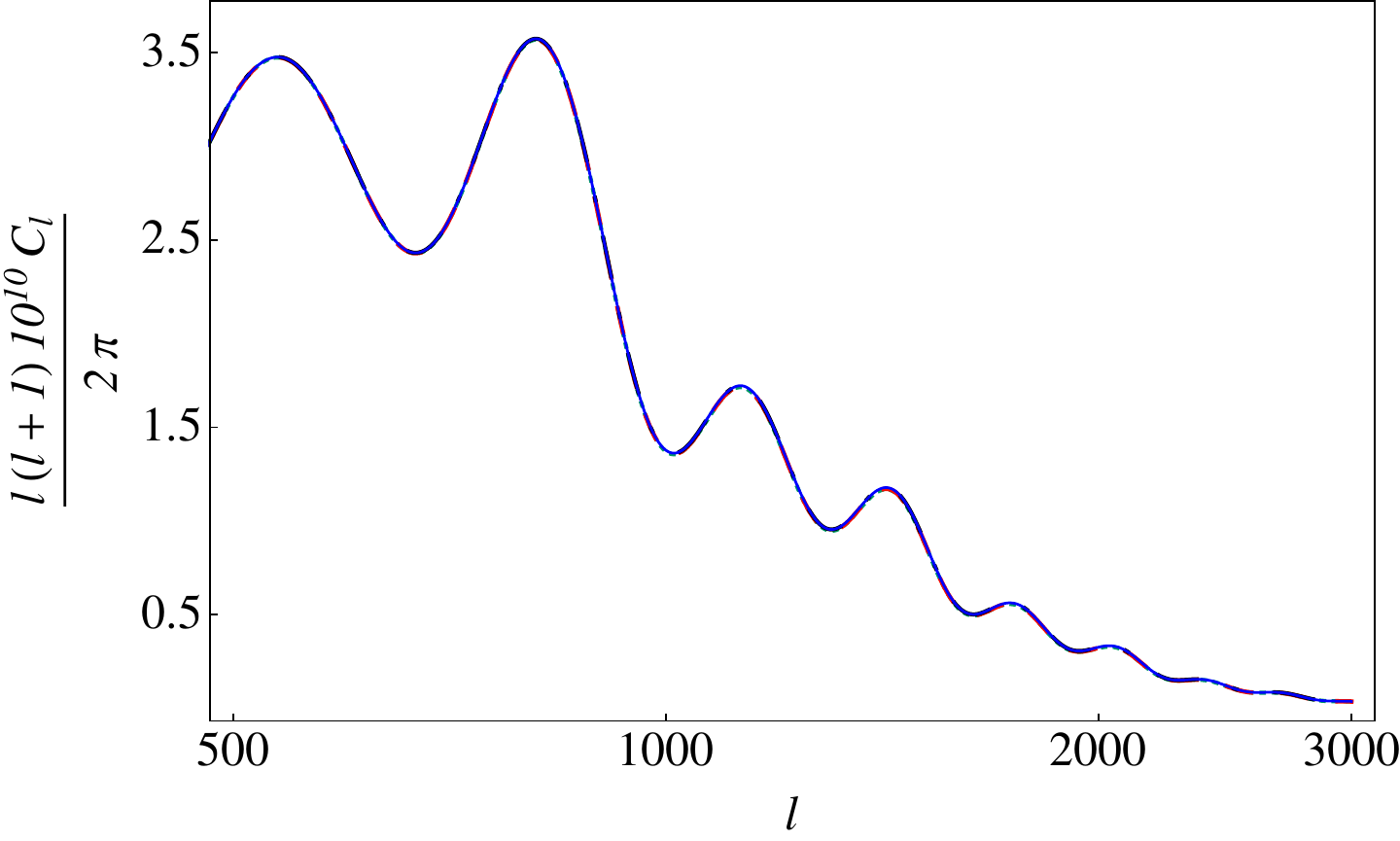}
\caption{\small Temperature spectrum of the CMB for low (left) and 
high (right) multipoles, computed for $r=0,~\alpha=0$ (blue); $r=0.16,~\alpha=0$ (black dot-dashed); $r=0,~\alpha=-0.021$ (green dotted) and $r=0.145,~\alpha=-0.015$ (red dashed). The other cosmological parameters are the same for the four curves and all primordial parameters correspond to $k_*=0.05$~Mpc$^{-1}$. In particular, $n_s=0.96$ and $A_s=2.3\times 10^{-9}$. A non-zero $r$ enhances the power at low $\ell$ and a negative $\alpha$ reduces it. The two parameters are degenerate and their effects can approximately cancel. For large $\ell$ the four curves overlap and there is no tension with the current CMB data by introducing $r$ or $\alpha$ at $k_*$. These spectra have been computed with CLASS \cite{Lesgourgues:2011re}.}
\end{figure}

While the situation becomes clarified, it is interesting to explore  the implications that an observation of primordial B-modes would have for cosmological inflation. The Planck data indicates a ``lack'' of power in the temperature power spectrum at low $\ell$ with respect to the best fit parameters \cite{Ade:2013zuv}. This deficit  was already present in WMAP data \cite{Hinshaw:2012aka}. If a detection of a sizable $r$ is achieved,  the total CMB power at low multipoles in the fit would  be enhanced, thus strengthening this deficit in the data. A value of $r\sim 0.1$, which cannot yet be excluded, would be just around the  upper bound of $r<0.11$ at $k=0.002$\,Mpc$^{-1}$, ($95\%$ C.L.) obtained by the Planck collaboration \cite{Ade:2013uln} by combining their own CMB temperature data with the large-$\ell$ data of SPT \cite{Hou:2012xq} and ACT \cite{Sievers:2013ica}, and polarization information from WMAP \cite{Hinshaw:2012aka}. The fit could then be improved by including a small scale dependence of the scalar spectral index $n_s$ \cite{Ade:2014xnap}, which would compensate the enhancement of power due to $r$  (see Figure \ref{cls}). Once the running of $n_s$ is taken into account, the Planck bound rises to $r<0.23$ at $k=0.05$\,Mpc$^{-1}$ ($95\%$ C.L.) with constant  $\alpha\equiv d\,n_s/d\ln k = -0.022^{+0.011}_{-0.010}$ at the same scale and confidence level \cite{Ade:2013uln}.\footnote{Taken at face value, the BICEP2 interpretation of their own data would be in tension with Planck. It has been argued that the respective 1-d marginalized posteriors for $r$ at $0.05$\,Mpc$^{-1}$, assuming $n_t=0$, overlap sufficiently to disregard the tension \cite{Audren:2014cea}. However, such qualitative estimate could be inappropriate for the comparison of two separate experiments. It would be more adequate for determining (in an experiment) the odds that a single event is compatible (at a certain confidence level) with an a priori probability distribution. A different estimation was done in \cite{Smith:2014kka} taking into account the deficit of power at large scales. According to it, the likelihood of the tension between Planck and BICEP2 assuming $\alpha=0$ would be of $\sim 1/1000$.} An analysis of BICEP2 and Planck which takes into account the recently released Planck polarization data \cite{Adam:2014bub} indicates $\alpha=-0.018\pm 0.009$ and $r<0.116$, both of them at $k_0=0.05$Mpc$^{-1}$ \cite{Cheng:2014pxa}.

It has often been claimed that a large running of the spectral index is incompatible with a long slow-roll inflationary period, see e.g. \cite{Easther:2006tv}. If this were the case, one would expect that a large value of $r$ would make even more difficult achieving enough inflation. In this work, we tackle this issue and analyse whether a negative running $\alpha\sim -{10^{-2}}$ and a substantial amount of primordial gravity waves $r\sim 10^{-1}$ are compatible with sufficient slow-roll inflation to solve the horizon problem.

Remarkably, a significant detection of primordial B-modes could discard most single field slow-roll inflation models.\footnote{In this work we focus on slow-roll inflation in which the whole primordial spectrum of perturbations is entirely produced by a single rolling field. Models involving more scalars that impact the spectrum e.g.\ capable of generating large negative $\alpha$ \cite{Sloth:2014sga} or a sizable $r$ \cite{Biagetti:2013kwa} are also possible.}  The precise determination of the scalar spectral index by Planck: $n_s\simeq 0.96$ at $k=0.05$\,Mpc$^{-1}$ (both with and without running) already selected just a handful of them \cite{Ade:2013uln}. Most of these models would turn to be excluded by a measurement of a large value of $r$.  The reason is related to the so called ``Lyth bound" \cite{Lyth:1984gv}. Given that at least $\sim 60$ e-folds are needed for the horizon problem, if we impose slow-roll and assume that $r$ is essentially constant during inflation, we get that
$\Delta\phi/M_P\sim 60\sqrt{r/8}$ is the minimal variation of the inflaton that is required. For example, $r=0.1$ gives $\Delta\phi\simeq 7\, M_P$, indicating a super-Planckian field excursion and selecting ``large-field'' models. The additional requirement of a sizable negative $\alpha$  would reduce further, and drastically, the set of acceptable models.\footnote{As we have discussed, a detection of $r\sim 0.1$ would worsen the problem of the deficit of power at small $\ell$ under the assumption of $\alpha=0$. In such a situation judging whether $\alpha$ is actually required by the data would typically be done using statistical methods and applying ``Occam's razor''. However, $\alpha\neq 0$ (and the same for higher order coefficients in the $n_s$ expansion) is a general prediction of single field inflation as it is also $n_s\neq 1$; and from this point of view it should be considered in a fit to the data.} In particular, the quadratic potential $V=m^2 \phi^2/2$, the simplest model able to yield a large amount of tensor modes (and enough inflation) would be excluded on these grounds, since it predicts too little running, $\alpha\simeq -r^2/32 \sim -3\times10^{-4}$ for $r\sim 0.1$. The same is true for most single field slow-roll models  capable of producing a value of $r$ of that order, e.g. natural inflation \cite{Freese:1990rb} and axion monodromy inflation \cite{McAllister:2008hb}. Therefore,  if the tensor-to-scalar ratio were determined to be $r\sim 0.1$, the requirement of a large $\alpha\sim -0.01$ as a natural way to accommodate the data would represent an important challenge for inflation model-building.

In this paper, after discussing the difficulties that appear by requiring sizable $r$ and $\alpha$ (Section 2), we present a general class of models (Section 3) that is able to overcome them, as explained in Section 4. Potentials of this type are capable of producing enough inflation ($N_e > 50$--$60$) within the slow-roll approximation (which we review in the Appendix). 

\section{Difficulties arising from large $r$ and $\alpha$}
If inflation is produced by a scalar field $\phi$ rolling slowly down a potential $V$, the primordial scalar and tensor power spectra can be computed in terms of the first few slow-roll parameters
\begin{align}\label{slowrolldef}
\epsilon =\frac{M_P^2}{2}\left(\frac{V'}{V}\right)^2\,,\quad \eta =M_P^2\frac{V''}{V}\,,\quad \xi = M_P^4\frac{V'V'''}{V^2}\,,
\end{align}
where $M_P$ is the reduced Planck mass $M_P=1/\sqrt{8\pi\,G}\simeq 2.435\times10^{18}\text{GeV}$ and $V'=d V/d\phi$. The scalar and tensor power spectra are parametrized as
\begin{align}
P_s(k)=  A_s\left(\frac{k}{k_*}\right)^{n_s-1+\frac{\alpha}{2}\ln\frac{k}{k_*}+\cdots}\,,\quad 
P_t(k) = A_t\left(\frac{k}{k_*}\right)^{n_t+\cdots}\,,
\end{align}
where all the parameters  are evaluated at some fiducial scale $k_*$, say $k_*\equiv0.05$\,Mpc$^{-1}$. At leading  order in the slow-roll expansion
\begin{align} \label{index}
n_s\simeq 1+ 2\,\eta-6\,\epsilon\,, \quad
\alpha\equiv \frac{d n_s}{d \ln k}\simeq -2\,\xi +16\,\epsilon\,\eta-24\,\epsilon^2\,.
\end{align}
and
\begin{align}
\label{Asth}
A_s\simeq \frac{1}{24\pi^2\epsilon}\frac{V}{M_P^4}\,,\quad A_t\simeq \frac{3}{2\pi^2}\frac{V}{M_P^4}\,.
\end{align}
Therefore, the tensor-to-scalar ratio is given by
\begin{align} \label{r}
r\equiv \frac{P_t}{P_s}\simeq \frac{A_t}{A_s}\simeq 16\,\epsilon\,,
\end{align}
while $n_t$ is entirely determined by $r$ in slow-roll inflation through the consistency relation $n_t\simeq-r/8$\,. 

We will now analyze the difficulties that appear from requiring ``large'' values of $r$ and  $\alpha$. For the sake of illustration, let us take the following values at $k_*$:
\begin{align} \label{nsg}
n_s\simeq 0.96\,,
\end{align}
as it is already rather well constrained by the data \cite{Ade:2013uln}, and
\begin{align} \label{rg} r \simeq 0.1\,,\quad \alpha\simeq -0.01\,,
\end{align} 
which translates into\footnote{Notice that the values of $\epsilon$ and $\eta$ are fixed by $r$ and $n_s$, while $\xi$ depends also on $\alpha$.}
\begin{align} \label{derigoal2}
\epsilon\ \simeq\ -5\eta\  \simeq\  \frac{4}{3}\xi\ \simeq\  6\times10^{-3}\,,
\end{align} 
meaning that the second derivative of the inflaton potential (in Planck units) must be much smaller than the first and third derivatives. To see this more clearly, we can normalize the potential and the inflaton as
 \begin{align}
\mathcal{V}=V/V_*\ , \;\;\;\; \varphi=\phi/M_P\,,
 \end{align}
where $V_*=V(\phi_*)$ with $\phi_*$ being the value of the inflaton at $k_*$. In these units,
\begin{align} \label{deris}
\mathcal{V}'=\sqrt{2\epsilon}\,, \quad
\mathcal{V}''=\eta\,,\quad \mathcal{V}'''=\xi/\sqrt{2\epsilon}\,,
\end{align}
where primes denote here the derivatives with respect to the normalized inflaton, $\varphi$. For instance, from \eq{derigoal2} we get 
$\mathcal{V}''\simeq -0.01\ \mathcal{V}'$ and $\mathcal{V}'''\simeq 0.36\ \mathcal{V}'$, with $ \mathcal{V}'\simeq 0.11$. This unusual hierarchy of derivatives,
\begin{align} \label{derigoalmod}
|\mathcal{V}'|\gg|\mathcal{V}''|\ll|\mathcal{V}'''|\,,
\end{align}
suggests a rather special shape of the potential and it is the first of the problems that arise from the requirement of large $r$ and large negative $\alpha$. For example, it is easily checked that a renormalizable potential of the form ${V} = \Lambda + \mu^2 \phi^2 + \lambda \phi^4$, cannot satisfy (\ref{derigoal2}).

Given the scalar spectral index and  the tensor-to-scalar ratio  at some scale $k_0$, we can get their values at another scale using the expansions
\begin{align} \label{scaling}
n_s(k)= n_s(k_0)+\alpha \ln \frac{k}{k_0}+\cdots\,,\quad r(k)= r(k_0)+\alpha_r \ln \frac{k}{k_0}+\cdots\,,
\end{align}
where the running of $r$ with the scale is given by
\begin{align} \label{rrun}
\alpha_r \simeq -r(n_s-1)-\frac{r^2}{8}\,,
\end{align}
which can be easily obtained by simply taking the derivative of \eq{r} with respect to $\ln k$ and using $d\phi/d\ln k \simeq -M_P \sqrt{2\epsilon}$. From these formulas it is straightforward to check that the variation of the primordial parameters from $k_*$ to a relatively close scale (e.g. $0.002$ Mpc$^{-1}$) is very small. Hence, the conclusion about the relative size of the derivatives does not depend substantially on the specific fiducial scale we choose (as far as it is reasonably close to the present Hubble scale). 

The second problem appears from the requirement of a sufficient number of e-folds. Since the fiducial scale $k_*=0.05$\,Mpc$^{-1}$ corresponds to approximately $4$ e-folds after the beginning of observable inflation, the inflationary model should provide at least $\sim 50$--55 e-folds from the fiducial point to the end of inflation.  In the slow-roll approximation the number of e-folds, $N_e$, can be expressed as 
\begin{align} \label{ne}
N_e \simeq  \int_{\varphi_e}^{\varphi_i} \frac{d\,\varphi}{\sqrt{2\epsilon}}\,,
\end{align}
where the $i$ and $e$ subscripts denote the initial and the final values of the field. Now, in addition to the fact that the ``large'' initial value of $\sqrt{2\epsilon}\sim 0.1$ will typically make it challenging to produce $\sim 55$ e-folds, this problem can be easily aggravated by the ``large'' second order slow-roll parameter $\xi$. Naively, a large $\xi$ may  trigger the break-down of slow-roll, and thus of inflation, too early.

To illustrate this, consider the expansion of the inflationary potential around the value of the field at the fiducial scale:
\begin{align} \label{Vexpanded}
\mathcal{V} = 1 + \sqrt{2\epsilon}(\varphi - \varphi_*)+ \frac{1}{2} \eta (\varphi-\varphi_*)^2 + \frac{1}{3!}  \frac{\xi}{\sqrt{2\epsilon}}(\varphi-\varphi_*)^3+\cdots\,,
\end{align}
where $\varphi_{*}=\phi_{*}/M_P$. Neglecting higher order terms (i.e.\ assuming that this is the shape of the potential during the whole period of inflation) and given the values (\ref{derigoal2}), this potential only produces approximately $15$--$18$ e-folds  between the fiducial scale and the end of inflation,\footnote{This can be checked solving \eq{evo} until the condition $\epsilon_H=1$ is reached. See Appendix.} which is far insufficient to solve the horizon problem. This shows that the potential must deviate from this shape at larger field displacements.

In summary, sizeable values of $\epsilon$ and $\xi$ require a quite peculiar hierarchy of derivatives, that cannot be obtained naturally within most of the physically meaningful models of inflation. In addition, the unusually large third derivative would typically make it challenging to avoid the breakdown of slow-roll too early during inflation.

Despite these two difficulties, we are going to present a model, which is effectively of single field slow-roll type, that can be naturally realized in models of particle physics and is able to fulfil the required conditions. This model, that we name {\it Instep Inflation}, is able to accommodate a large running and a large tensor-to-scalar ratio, while still producing enough inflation. The model does not rely on an exotic or ad-hoc potential, non-smooth ``features'' or a breaking of slow-roll.

\section{Instep Inflation }
In broad classes of particle physics models, the scalar potential has flat directions at tree-level, e.g. from the presence of an accidental symmetry in the scalar sector. Flat directions are also typical in supersymmetric and string theories, which have large moduli spaces at tree-level without the need of any symmetry reason behind. We will now see how flat directions allow to solve the difficulties discussed in the previous section.

The amplitude of primordial scalar perturbations given by Planck is $\ln(10^{10} A_s)=3.100\pm0.030$ at $k=0.002$\,Mpc$^{-1}$ ($95\%$ C.L.) \cite{Ade:2013uln}. This corresponds to 
\begin{align}
\label{As}
A_s=(1.75\pm 0.05)\times 10^{-9}\  (95\%\  {\rm C.L.})\ {\rm at}\  k_*=0.05\ {\rm Mpc}^{-1}\,.
\end{align}
Then, according to eq.~(\ref{Asth}), 
 \begin{align} \label{altura}
V \simeq  \frac{r}{0.1}\left(1.4\times 10^{16} \text{GeV}\right)^4 \,,
 \end{align}
which for $r\sim 0.1$ is remarkably close to the supersymmetric grand unification scale, providing further motivation for such scenarios in SUSY gauge-unification models. Actually, the class of models that we will examine now can naturally  (but not necessarily) live in the realm of supersymmetric theories.

Since flat directions in the tree-level potential are usually not protected by any exact symmetry, they are generically lifted by radiative corrections. In some cases the lifting is only logarithmic
with a one-loop potential of the form $\Delta V_1 \sim \log (m^2/Q^2)$, where $m^2$ is some mass-squared eigenvalue that depends on the field (or fields) along the flat direction
and $Q$ is the renormalization scale. These cases provide natural candidates for slow-roll directions and have been used in the past to construct models of hybrid type (e.g. \cite{Binetruy:1996xj}). In these models, the rolling along the lifted flat direction ends when the inflaton field reaches a critical value, triggering a VEV for a second field (the waterfall field) that deflects it from the flat direction.\footnote{However, these models (at least in their original formulations \cite{Linde:1993cn,Binetruy:1996xj}) would be in conflict with the detection of a sizable tensor-to-scalar ratio.}

On the other hand, one generically expects new physics at scales $\Lambda$ beyond the $10^{16}$ GeV scale at which inflation takes place. Such new physics could typically affect the inflaton potential
via non-renormalizable operators supressed by powers of  $\Lambda$. 
It is then natural to have inflationary potentials based on flat directions  lifted both by radiative corrections (LOG lifting) and from non-renormalizable operators (NRO lifting) arising from physics beyond the inflatonary scale. Writing the dominant contributions of each type, we expect an inflaton potential of the form
\begin{align} \label{pot}
V(\phi)=\rho+\beta \log\left[\frac{m(\phi)}{Q}\right]+ g^{2N+2} \phi^4\left(\frac{\phi^{2}}{\Lambda^{2}}\right)^N\,.
\end{align}
Here $\rho$ is the tree-level ($\phi-$independent) potential before lifting. Since $\rho$ is roughly the energy density at the fiducial scale, it is constrained to be $\rho\simeq (1.4\times 10^{16}\ {\rm GeV})^4$ for $r\sim 0.1$. 

The radiative lifting is controlled by $m(\phi)$, which is the most relevant mass contributing to the radiative corrections; and $Q$, the renormalization scale. Typically, one expects $m(\phi)$ to have both field-independent and field-dependent contributions, e.g. 
\begin{align} 
\label{gmass}
m^2(\phi)=M^2+\kappa^2\phi^2\,,
\end{align}
where $M$ does not depend on $\phi$ and $\kappa$ is a dimensionless coupling.\footnote{As mentioned earlier, other dependences, such as $m^2(\phi)=(M+\kappa\phi)^2$ are common as well. It is easy to construct supersymmetric models which implement the two types of dependence. 
}
The $\beta$ parameter is the one-loop coefficient of the $\beta-$function of $\rho$, i.e.\ $\beta=d\,\rho/d\log Q$, as it follows from the invariance of the effective potential with respect to the renormalization scale, $Q$, at one-loop. For numerical work we will choose $Q=m(\phi_*)$, where $\phi_*$ is the value of the inflaton field at the fiducial scale (for practical purposes, the initial value of the inflaton) and thus the parameters of the potential are to be understood at this scale. This choice corresponds to a one-loop leading-log approximation of the potential.\footnote{Resummation of leading-logs to all orders corresponds to $Q=m(\phi)$, so that the potentially large logs dissapear from the explicit expression for $V$, and their $\phi-$dependence is transferred to $\rho=\rho(Q=m(\phi))$, with $\rho$ evolved by integrating its one-loop beta function. Our one-loop expression will be a good approximation to the full potential if the running of $\beta$ itself is sufficiently small\cite{Ballesteros:2005eg}.}

On the other hand, the NRO lifting is controlled by the scale of new physics beyond the inflationary energies, $\Lambda$; the degree of the most relevant NR operator, $4+2N$ (assumed to be even); and the dimensionless coupling $g$. The exponent of this coupling in eq.~(\ref{pot}) corresponds to a tree-level generated NR operator.
If the NR operator is instead generated at one loop, an extra factor $g^2/(16\pi^2)$ should be included.

In general, several log-terms, associated with different masses $m(\phi)$, and several NROs of different orders will be present in the potential (\ref{pot}). Here we only consider the most relevant contribution of each type. In the case of the NRO, an operator can be the dominant one if its coupling $g$ is sufficiently large or there is some symmetry reason that forbids terms with smaller $N$.

The class of inflationary potentials (\ref{pot}) was introduced in \cite{Ballesteros:2005eg} to explain the WMAP3 indication of a  negative $\alpha$ \cite{Peiris:2003ff}.\footnote{See \cite{Ballesteros:2007te} for a detailed Bayesian statistical analysis of the performance of a variant of these models (with $M=0$ in  \eq{gmass}) using WMAP and a set of LSS data. The advantage of using the precise predictions of the models for the primordial spectrum from an integration of the slow-roll expansion equations was emphasized there.}  It was shown there  that for $M=0$ such potential is able to produce  enough inflation and a large negative running of the scalar spectral index ($\alpha\sim -0.05$ at $k=0.002$ Mpc$^{-1}$) within the slow-roll approximation. However, the value of $r$ obtained in that case was negligible ($r\sim 10^{-3}$). 
We will now see that for $M\neq 0$ this kind of potential can provide a large tensor-to-scalar ratio ($r\sim 0.1$) and a significant running ($\alpha\sim -0.01$), without compromising the slow-roll approximation.
 
We end this section by counting the number of independent parameters of our potential \eq{pot}. Four of them are continuous: $\rho$, $\beta$,  $M/\kappa$ and $\Lambda/g^{1+1/N}$,
and one is discrete:\ $N$. As mentioned above, the value of $r$ fixes $\rho$. For consistency of our effective field theory approach, the inequality $\Lambda\gg m(\phi_*)$ should hold.

\section{Large $r$ and $\alpha$ with Instep Inflation}

\renewcommand{\eq}[1]{(\ref{#1})}

The first three slow-roll parameters of Instep inflation read
\begin{align} \label{deriv1}
\frac{V_*}{M_P}\sqrt{2\epsilon} & \simeq  
\frac{\kappa^2\beta \phi }{m^2(\phi)}\ +\ 2 (2+N) g^{2N+2}\phi ^3 \left(\frac{\phi }{\Lambda }\right)^{2 N}\ ,\\ \label{deriv2}
\frac{V_*}{M_P^2}\eta& \simeq \kappa^2\beta  \frac{(M^2-\kappa^2\phi^2)}{m^4(\phi)}\ +\ 2 (2+N) (3+2 N) g^{2N+2}\phi ^2 \left(\frac{\phi }{\Lambda }\right)^{2 N}\ ,\\ \label{deriv3}
\frac{V_*}{M_P^3}\frac{\xi}{\sqrt{2\epsilon}}& \simeq 2\kappa^4\beta\phi  \frac{ \left(-3 M^2+\kappa^2\phi ^2\right)}{m^6(\phi)}\ 
+\ 4 (1+N) (2+N) (3+2 N) g^{2N+2}\phi  \left(\frac{\phi }{\Lambda }\right)^{2 N}\,,
\end{align}
where $V_*=V(\phi_*)\simeq \rho$. It is easy to see that neither the LOG nor the NRO pieces of these expressions have separately the required forms to fulfil the pattern of eq.~(\ref{derigoalmod}), in particular the hierarchy $\eta^2/\xi \sim {\cal O}(10^{-3})$. If we consider only the LOG contribution, and demand $\xi>0$ to get a negative $\alpha$, we find $\eta^2/\xi>1/2$. Similarly, if we focus on the NRO alone it turns out that $\eta^2/\xi=(3+2N)/(2+2N)\sim\mathcal{O}(1)$. This means that  if a sizable $r\sim 0.1$ is measured {\it and} a large negative $\alpha\sim -0.01$ is significantly required by the data, the discrepancy with respect to the LOG and the NRO predictions would rule out ``loop-inflation'' models (see e.g. \cite{Martin:2013tda})  and any ``chaotic'' monomial model of inflation. Note that these difficulties arise even before starting to address the issue of obtaining a sufficient number of e-folds and illustrate further the challenge of finding suitable inflationary scenarios with $r\sim 0.1$ and $\alpha\sim -0.01$. 

The eqs.~(\ref{deriv1}--\ref{deriv3}) show that a rather obvious way to obtain the desired structure of eq.~(\ref{derigoalmod}) is to consider both contributions (LOG and NRO) simultaneously. The LOG contribution is the dominant one for small $\phi$, whereas the NRO becomes the most important at large $\phi$, so there is a range of the inflaton walk  for which the LOG and NRO pieces have comparable sizes. Given the negative sign of the LOG contribution to the second derivative of the potential at sufficiently large $\phi$  (see eq.~(\ref{deriv2})), a partial cancellation between the LOG and NRO contributions to $\eta$ takes place. Hence, it is the interplay between the two contributions that allows  ${\mathcal V}''$  to be suppressed with respect to ${\mathcal V}'$ and ${\mathcal V}'''$, as required. It is important to emphasize that both types of corrections to the flat direction are expected in general and therefore it is sensible to consider them together. 

 \begin{table}[bt] 
\begin{center}
\begin{tabular}{l*{7}{c}r}
$N$   &$M/(\kappa M_P)$  & $\beta/\rho$  & $\Lambda/M_P$ & $\phi_*/M_P$  & $g$ & $N_e$ & $\alpha$  \\
\hline
1 & 1.10 & 0.34 & 7.67 & 3.09 & $9\times 10^{-4}$  & 55 & -0.01\\
1 & 0.80 & 0.23 & 13.95 & 2.06 & $2\times10^{-3}$ & 60 & -0.02\\
2 & 1.50 & 0.41 & 7.28 & 3.29 & $1\times 10^{-2}$  & 95 & -0.01\\
2 & 0.93 & 0.29 & $12.36$ & 2.51 & $2\times 10^{-2}$   & 60 & -0.02\\
4 & 1.70 & 0.55 & $11.27$ & 4.56 & $8\times 10^{-2}$  & 60 & -0.01\\
5 & 2.00 & 0.59 & $8.06$ & 4.71 & $1\times 10^{-1}$   & 80 & -0.01\\
6 & 2.20 & 0.64 & $12.33$ & 4.96 & $2\times 10^{-1}$  & 80 & -0.01\\
\end{tabular}
\caption{\small Examples of parameters of the potential \eq{pot} that give $ n_{s}= 0.96$, $r= 0.1$ and $\alpha$ as indicated in the last column. For these values we have taken $A_s=2\times 10^{-9}$, which implies that the energy scale of the potential is fixed to be $\rho\simeq (1.465\times 10^{16}\ {\rm GeV})^4$ by the value of $r=0.1$. The renormalization scale $Q$ is set to $m(\phi_*)$. Notice that $g$ and $\Lambda$ combine into a single parameter: $\Lambda/ g^{1+1/N}$, though we have taken some reasonable values for $g$ for the sake of illustration. The column $N_e$ gives approximately the number of e-folds before the inflaton reaches the minimum of the potential.} \label{table1}
\end{center}
\end{table}

Let us illustrate now how the potential \eq{pot}, with particular choices of the parameters, can reproduce satisfactorily the pattern of derivatives (\ref{derigoalmod}). Table 1 shows some working examples. Actually, there are many acceptable possibilities for reasonable values of the parameters.\footnote{An in-depth exploration of the parameter space and a Bayesian statistical analysis of the performance of the model  could be performed using a Monte Carlo method in combination with a tool such as ModeCode \cite{Mortonson:2010er}.} In Figure 2 we show the shape of the potential for three such examples.\footnote{Incidentally, the name ``Instep inflation" was motivated by this shape.}

\begin{figure}[bt] \label{pots}
\centering
\includegraphics[width=0.60\textwidth]{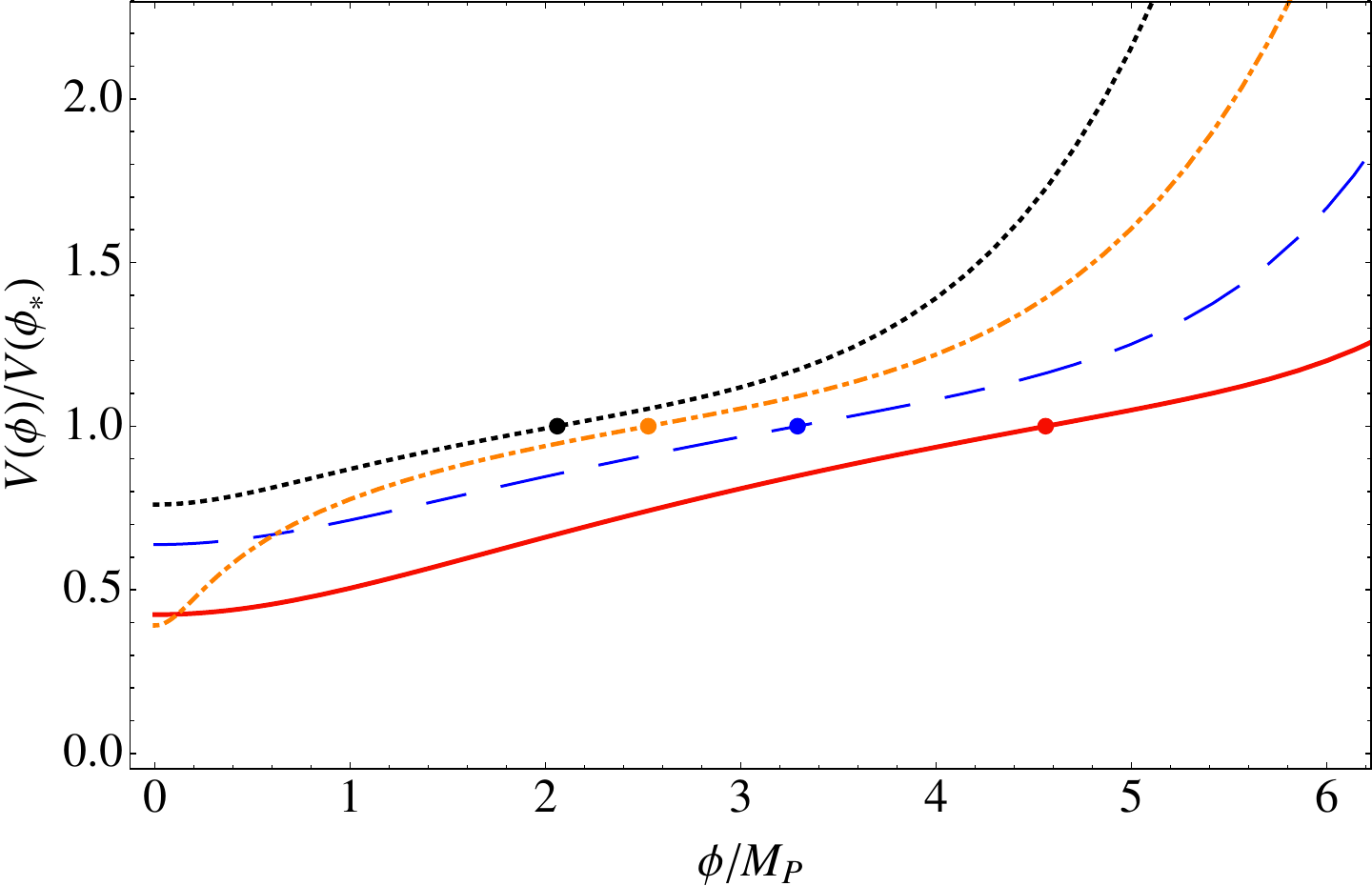}\quad
\caption{\small Instep inflationary potentials from Table \ref{table1}. $N=1$, $M=0.80M_P$ (black dotted); $N=2$, $M=1.50M_P$ (blue dashed) and $N=4$, $M=1.70M_P$ (red continuous). The potentials have been normalized at $\phi_*$, shown for each case with a disk. This is the point corresponding to $k_*=0.05$\,Mpc$^{-1}$, at which we compute the values of the primordial observables and start the numerical integration of the inflaton trajectories. For comparison we also show an example with $N=1$, $M=0.20M_P$ (orange dash-dotted) that gives $n_s=0.96$, $r=0.1$ and $\alpha=-0.02$, but only produces 4.4 e-folds before the breaking of slow-roll inflation.}
\end{figure}

We close this section with some comments concerning the slow-roll approximation and the expansion in slow-roll parameters, the computation of the number of e-folds and the end of inflation. 

\vspace{0.2cm}

\noindent {\bf I. Slow-roll approximation.}

It is often claimed that a ``large'' running of the scalar spectral index implies the premature breakdown of slow-roll and an inflaton potential too steep to sustain inflation for more than a few e-folds $(N_e \ll 60)$. Certainly, it could be naively expected that a non-negligible $\xi$ meant that  higher order derivatives are expected to be important at all times, stopping inflation soon after it starts. However, this is not necessarily the case.\footnote{For instance, the conclusion of \cite{Easther:2006tv} that a large $\alpha$ implies the breakdown of slow-roll before producing 30 e-folds, relies implicitly on  a third order Taylor expansion of the potential as in eq.~\eq{Vexpanded}. Such an expansion can easily fail to describe the actual shape of the potential just a few e-folds away from $\varphi_*=\varphi(k_*)$. Therefore, a value $\alpha\lesssim -0.02$ at given $k_*$ does not generally prevent from reaching 60 e-folds while still keeping the whole trajectory of the inflaton in the slow-roll regime.} The slow-roll approximation is valid if and only if the Hubble slow-roll parameters (see eqs.~(\ref{eH}, \ref{eH2}) of the Appendix) satisfy $\epsilon_H\ll 1 $ and $|\eta_H|\ll 3$. These parameters are related to the potential slow-roll parameters, $\epsilon$ and $\eta$, by the system of differential equations (\ref{exact1}) \cite{Liddle:1994dx}.  In order to solve this system, an initial condition for the field velocity needs to be supplied, so the values of $\epsilon$ and $\eta$ at a given point are not enough to ascertain or disprove the validity of the slow-roll approximation. The need of specifying the initial conditions cannot be overcome by knowing $\xi$ or any higher order potential slow-roll parameters, which only give purely geometrical (and not dynamical) information. The only way of reaching a conclusive statement is solving the dynamics of the inflaton.

As it is well known, the slow-roll attractor solution will be quickly reached provided that the initial velocity of the inflaton is not too different from the attractor one, given in \eq{att}. We have confirmed that this is indeed what occurs for all the examples in Table 1 by integrating the equations of motion for a wide range of initial conditions, checking that $\epsilon_H\ll 1 $ and $|\eta_H|\ll 3$ hold true during the entire trajectory.
\newline

\noindent {\bf II. Slow-roll expansion.}

The fact that the slow-roll approximation holds does not ensure that we can use eqs.~(\ref{index}) and (\ref{r}) to accurately compute $r$, $n_s$, and $\alpha$, close to the beginning  of inflation. In principle, higher order parameters could be necessary.\footnote{See e.g. \cite{Liddle:1994dx} for their definition and their effect on $r$, $n_s$, and $\alpha$, etc.} We have computed several of these higher order slow-roll (both conventional and Hubble) parameters, checking that their contribution to $r$, $n_s$, and $\alpha$ is negligible for the examples of Table \ref{table1} at the beginning of inflation.\footnote{The fact that higher order slow-roll parameters are negligible ensures that neglecting the running of the running of the scalar spectral index is a good approximation as well.} Therefore, we have not only checked that the slow-roll {\it approximation} is satisfied, but also that the slow-roll {\it expansion} is. 
\newline

\noindent{\bf III. Number of e-folds.} \\
Inflation goes on as long as $\epsilon_H<1$, so  we should compute the number of e-folds integrating $H$ over time until that condition is achieved. Without doing any approximation, the number of e-folds between some initial time $t_i$ and the end of inflation at $t_e$ is 
\begin{align} \label{neH}
N_e=\int_{t_i}^{t_e} H\, dt\,.
\end{align}
In practice, we solve the dynamics of the inflaton using $N_e$ as time variable, see eq.~\eq{evo} in the Appendix. This allows to get very easily the number of e-folds that are produced after the field goes past a given point.

It is easy to find choices of the parameters of our Instep potential \eq{pot} that give \eq{derigoal2} and for which $\epsilon_H$ reaches 1 (and keeps growing) after a certain point. The total number of e-folds in those cases is typically small: $N_e\sim 15$ to $20$. However, for the examples of Table \ref{table1} (and many others),  $\epsilon_H$ grows from its small initial value, reaches a  maximum ($<1$) and then decreases again, while the inflaton approaches $\phi=0$. The number $N_e$ indicated in Table 1 corresponds approximately to the amount of inflation produced from $\phi=\phi_*$ (where we compute the values of the primordial parameters) and some $\phi\simeq 0$, close to the minimum of the potential. All of those examples produce sufficient e-folds to solve the horizon problem.
\newline

\noindent{\bf IV. End of inflation.}\\
The minimum of the potential at $\phi=0$ corresponds to a value $V(0)\neq 0$ and larger than the measured value of the cosmological constant, so a mechanism that eventually ends inflation is required. This should not occur earlier than $\sim$50 e-folds after $\phi=\phi_*$\,. As mentioned in Section 3, this mechanism can be provided by a waterfall field, as in standard hybrid models \cite{Linde:1993cn}. Actually, for small $\phi$, i.e.\ at the last stages of inflation, the NRO contribution to the inflaton potential (\ref{pot}) is completely negligible. Consequently, the dynamics of the end of inflation can be exactly as in many hybrid inflation models with flat directions, where only the LOG contribution is present. When the inflaton is near the value $\phi=0$, the waterfall gets destabilized, rolls down to the absolute minimum at $V\simeq 0$ and inflation ends. In most of these cases (e.g. D-term inflation \cite{Binetruy:1996xj}), the inflationary period is correctly described using a single field, as we have done in this paper. The waterfall only becomes relevant when $\phi$ approaches its minimum.  In this sense the model is of single-field type, since both the expansion and the primordial spectrum are generated by the same field.

\section{Conclusions}

If a value of the tensor-to-scalar ratio, $r\sim{\cal O}(10^{-1})$, is eventually established from CMB observations, a large fraction of  single field slow-roll models will be ruled out. Those that will survive generically exhibit a super-Planckian field excursion. If, in addition, it turns out that a substantial running of the scalar spectral index, $\alpha\sim -{\cal O}(10^{-2})$, is significantly required by the data, the cut in the set of allowed models will be severe. There are very few proposals in the literature attempting to explain these features simultaneously within a well motivated scenario from the point of view of particle physics.\footnote{A non-Bunch-Davies vacuum may help to obtain both a negative running and a sizable tensor-to-scalar ratio \cite{Ashoorioon:2014nta}.}\footnote{A finely tuned fifth order polynomial may seemingly produce large $r$ and $\alpha$ at large scales \cite{BenDayan:2009kv}.  However, the slow-roll expansion is generally not valid in that case and the running is strongly scale dependent, even up to very small scales. 
}\footnote{There exist some single-field models that share with Instep Inflation the property of having an asymmetric tilted plateau. One such example is Fibre Inflation \cite{Cicoli:2008gp}. In slow-roll this model predicts $r\simeq 0.005-0.01$. However, it might lead to $r\simeq 0.1$ and enough suppression of power at low multipoles, provided that slow-roll breaks down around the inflection point of the potential \cite{Cicoli:2013oba,Pedro:2013pba}. Another example, arising from supergravity \cite{Kallosh:2014xwa}, is the potential $V(\phi)=m^2\phi^2\left(1-a\, \phi+a^2\, b\,\phi^2\right)^2/2$, where $m^2$, $a$ and $b$ are real numbers. The value of $m^2$ is fixed by the amplitude of primordial perturbations. For appropriate values of the parameters, the model can reproduce $n_s=0.96$ and $r\simeq 0.1$, together with a sufficient number of e-folds. This model might also reproduce the required suppression at low$-\ell$. For that goal, a non-minimal coupling of the inflaton to gravity would be required \cite{Kallosh:2014xwa}.
} 

Two are the main difficulties that arise from large $r$ and $\alpha$. Firstly, the shape of the inflationary potential requires a special pattern of derivatives that cannot be achieved with the simplest models e.g. renormalizable polynomials of even-degree, monomial potentials of any order, sinusoidal functions or pure radiative lifting. In particular, the second derivative of the potential (in Planck units) should be much smaller than the first and the third and subsequent ones.  Secondly, the presence of a large third derivative could be expected to produce a quick end of the inflationary process (well before $\sim 60$ e-folds are  produced). The production of a sufficient number of e-folds may also be hindered by the large value of $\epsilon$ required to fit a large $r$, see eq.~(\ref{ne}). This is indeed the case for the aforementioned examples, but it is not necessarily so for more general shapes of the inflaton potential.

We have shown that if sizable values of $r$ and $\alpha$ are found in the data, {\it Instep Inflation} can explain them and produce enough e-folds in the slow-roll regime. The model is based on the lifting of a generic flat direction by two kinds of physical effects. The first one is a logarithmic contribution arising from the radiative corrections to the inflaton effective potential. The second one is an extra lifting, at the higher end of field values, due to effective imprints from physics at a higher energy scale. These imprints appear  typically in the form of non-renormalizable operators in the effective theory. For suitable choices of the parameters of the model, the combination of the two effects provides the challenging pattern of derivatives (\ref{derigoalmod}) that is required, i.e.\ the suppression of the second derivative of the inflaton potential. In addition, the flattening effect of the logarithmic part of the potential, which is the relevant one for small values of the inflaton, ensures that enough e-folds can be produced before inflation terminates. 

A value of $r\simeq 0.1$ would imply that the energy scale of inflation is about $10^{16}$GeV, which is remarkably close to the supersymmetric grand unification scale. This adds  further motivation for these models of inflation based in flat directions, since the latter are very common in supersymmetric scenarios. 

Instep Inflation is effectively a single-field model since the entirety of the inflationary trajectory is described by the dynamics of a single field, which is also responsible for generating the primordial spectrum of perturbations. The end of inflation should occur by the action of a waterfall field through a hybrid mechanism when its VEV is triggered by the inflaton field approaching the minimum of the flat direction. In this respect Instep Inflation is like many other hybrid models based on flat directions. Hybrid inflation can be embedded rather easily in a supersymmetric framework, which makes it particularly appealing. In fact, popular models of hybrid inflation (e.g. \cite{Binetruy:1996xj}) are based on supersymmetric flat directions lifted by radiative corrections, which is exactly our effective context when the inflaton is at small field values.

The BICEP2 claim of a detection of primordial gravitational waves \cite{Ade:2014xnap} has turned out to be premature since their measurement of CMB B-modes could be entirely due to Galactic dust foregrounds. However, there is still room left for a detection of $r$ around or somewhat smaller than 0.1. Should such an important tensor-to-scalar ratio be finally found in the data, it is quite possible that a running of the scalar spectral index may also be needed, due to the current lack of power in the temperature spectrum at low multipoles. Finding viable slow-roll models of inflation in agreement with these features is rather challenging. To deal with this potential problem, we have proposed here a possibility that has a good motivation from the point of view of particle physics. 

 \section{Appendix: the slow-roll approximation}
In this Appendix we review  the slow-roll approximation and provide some  useful expressions for our analysis. We consider a scalar field $\phi$ in General Relativity with a standard kinetic term and a potential $V(\phi)$. We use dots to denote time derivatives and primes for derivatives with respect to $\phi$. Assuming a FLRW metric, the equation of motion of the field is 
\begin{align} \label{dyn}
\ddot\phi+3H\dot\phi+V'=0\,,
\end{align}
where the Hubble parameter $H=\dot a/ a$ is given by the Einstein equation
\begin{align} \label{fried}
3\,M_P^2 H^2=\frac{\dot\phi^2}{2}+V\,,
\end{align}
being $M_P$ the reduced Planck mass, $M_P=1/\sqrt{8\pi G}$, and $G$ the Newton's constant. 

By definition, the slow-roll approximation applies if \eq{fried} and \eq{dyn} can be respectively  approximated by
\begin{align} \label{sr1}
V \simeq &\,\, 3\,M_P^2 H^2  \\ \label{sr2}
V'\simeq & - 3H\dot\phi\,.
\end{align}
In this case, 
\begin{align} \label{att}
\dot\phi\simeq-\sqrt{2\epsilon V/3}\,,
\end{align} 
which is the slow-roll attractor velocity and where $\epsilon$ is defined in \eq{slowrolldef}.

Without loss of generality, we assume that the field rolls from larger to smaller values. This is the appropriate choice for the potential \eq{pot} studied in this paper and the reason for the minus sign in the previous expression for $\dot\phi$\,. 

The validity of \eq{sr1} and \eq{sr2} is controlled by the Hubble slow-roll parameters (HsrP) $\epsilon_H$ and $\eta_H$ that are defined as (see e.g. \cite{Liddle:1994dx}):
\begin{align}  \label{eH}
\epsilon_H\equiv 3\,\frac{\dot\phi^2}{\dot\phi^2+2V}\,, \quad \eta_H\equiv -\frac{\ddot\phi}{H\dot\phi}\,.
\end{align}
Therefore, \eq{sr1} is true if and only if $\epsilon_H\ll 1 $; and \eq{sr2} is true if and only if $|\eta_H|\ll 3$. This is the slow-roll approximation.

Notice that the definition of inflation as accelerated expansion, i.e.\ $\ddot a >0$, is equivalent to $\epsilon_H<1$. This is a purely kinematic condition and does not require any assumption about what causes the expansion. Alternatively, we can define $\epsilon_H\equiv-\dot H/H^2$  or equivalently  $\ddot a=aH^2\left(1-\epsilon_H\right)$. Note that $\epsilon_H$ can only be interpreted a {\it slow-roll} parameter once we assume that inflation is driven by a scalar field. Then, $\epsilon_H<1$ is equivalent to $\dot\phi^2<V$.

It is possible to define the HsrP in a way that resembles \eq{slowrolldef}, using the inflaton as time variable \cite{Liddle:1994dx}. Differentiating \eq{fried} with respect to time and plugging \eq{dyn} into the result, we obtain $2M_P^2\dot H+\dot\phi^2=0$. Then, writing $\dot H= H'\dot\phi$ we arrive to $\dot\phi=-2M_P^2H'$. We can now replace $\dot\phi$ by this expression in \eq{fried} to get $3\,M_P^2 H^2=2\,M_P^4\left(H'\right)^2+V$. Given a potential $V(\phi)$ and an initial condition $H_i=H(\phi_i)$, we can solve the previous equation to obtain $H$ as a function of $\phi$. Then, we can write\footnote{Notice that we use a different definition of $\xi_H$ than the one in ref. \cite{Liddle:1994dx}. Ours is related to theirs by a square root.}
\begin{align} \label{eH2}
\epsilon_H\equiv 2 M_P^2\left(\frac{H'}{H}\right)^2\,, \quad \eta_H\equiv 2 M_P^2\frac{H''}{H}\,,\quad \xi_H=4 M_P^4\frac{H' H'''}{H^2}\,,\quad\ldots
\end{align}

It is convenient to replace the time variable by the number of e-folds, 
\begin{align}
N_e(t)    =\int^{t}_{t_i} H {d \tilde t}\,,
\end{align}
where $t_i$ is the time at which inflation starts. Using the relation $\dot\phi=H\,d\phi/dN_e$ and applying the equation \eq{fried} to eliminate $H^2$ in \eq{dyn} via
\begin{align} \label{h2}
V=H^2\left(3 M_P^2-\frac{1}{2}\left(\frac{d\phi}{d N_e}\right)^{2}\right)
\end{align}
we obtain:
\begin{align} \label{evo}
\frac{d^2\phi}{dN_e^2}+3\,\frac{d\phi}{dN_e}-\frac{1}{2M_P^2}\left(\frac{d\phi}{dN_e}\right)^3+\left(3M_P-\frac{1}{2M_P}\left(\frac{d\phi}{d N_e}\right)^{2}\right)\sqrt{2\epsilon}=0\,,
\end{align}
which is a non-linear second order  differential equation for $\phi$. The end of inflation is marked by the point where  $\sqrt{V}=-H\,d\phi/dN_e$. Having obtained $\phi(N_e)$, we can use \eq{h2} to compute the HsrP and determine  whether the slow-roll approximation ($\epsilon_H \ll 1$ and $|\eta_H|\ll 3$) is valid at any given moment of the inflationary process. We can express the first HsrP concisely as
\begin{align} 
\epsilon_H=\frac{1}{2M_P^2}\left(\frac{d\phi}{d N_e}\right)^{2}\,,\quad \quad \eta_H=\epsilon_H-\left(\frac{d\phi}{dN_e}\right)^{-1}\frac{d^2\phi}{dN_e^2}\,.
\end{align}
Actually, it is possible to write a recursive succession for the derivatives of the inflaton with respect to $N_e$ that we can use to express the entire HsrP hierarchy: $\epsilon_H$, $\eta_H$, $\xi_H$, etc.
\begin{align} \label{der1}
\frac{d\phi}{dN_e}=-M_P\sqrt{2\epsilon_H}\,, \quad
\frac{d^2\phi}{dN_e^2}=\frac{d\phi}{dN_e}\left(\epsilon_H-\eta_H\right)\,,\quad
\frac{d^3\phi}{dN_e^3}=\frac{d\phi}{dN_e}\left(3\epsilon_H^2+\eta_H^2-5\epsilon_H\eta_H+\xi_H\right)\,, \ldots
\end{align}
Using these expressions and the evolution equation \eq{evo} we get the system of equations \cite{Liddle:1994dx}:
\begin{align}
\label{exact1}
\frac{\eta_H-3}{\epsilon_H-3}\sqrt{\epsilon_H}=\sqrt{\epsilon}\,,\quad
\sqrt{2\epsilon_H}\,\frac{\eta_H'}{3-\epsilon_H}\,+\left(\frac{3-\eta_H}{3-\epsilon_H}\right)\left(\epsilon_H+\eta_H\right)=\eta\,,
\end{align}
where the second of them is equivalent to
\begin{align}
\xi_H  = 3(\eta_H-\eta)-\eta_H^2+\epsilon_H\left(3+\eta\right)\,.
\end{align}
At any given order, the relationship between {\it potential} ($\epsilon$, $\eta$,\,\ldots) and {\it Hubble} ($\epsilon_H$, $\eta_H$,\, \ldots) slow-roll parameters is differential rather than algebraic. Concretely, the $n-$th order HsrP can be written as a function of $\epsilon_H$ and the potential slow-roll parameters (PsrP) up to order $n-1$. In the slow-roll attractor regime \eq{att}, we have $\epsilon_H\simeq \epsilon$ and $d \phi / d N_e\simeq M_P\sqrt{2\epsilon}$. Therefore, only if the slow-roll approximation is valid the dynamical information is contained in the shape of $V(\phi)$; and in that case all the HsrP are entirely determined by the PsrP. 

\section*{Acknowledgments} \label{ackn}
We are indebted to J. R. Espinosa for his generous collaboration in relevant parts of the paper and
proposing the name {\it Instep Inflation}. We also thank B. Audren, T. Tram and M. Tucci for interesting exchanges.
This work has been supported by DFG through the project TRR33 ``The Dark Universe'',  the MICINN (Spain) under contract FPA2010- 
17747; Consolider-Ingenio CPAN CSD2007-00042, as well as MULTIDARK CSD2009- 
00064. We thank as well the Comunidad de Madrid (Proyecto HEPHACOS 
S2009/ESP-1473) and the European Commission (contract PITN-GA-2009-237920). 
Finally, we acknowledge the support of the Spanish MINECO’s “Centro de  
Excelencia Severo Ochoa” Programme under grant SEV-2012-0249.

\bibliographystyle{unsrt}
\begin{small}

\end{small}
\end{document}